\documentclass{ws-ijgmmp}
\def\draftnote{}
\usepackage{amsmath}   
\usepackage{amssymb}   
\usepackage{bm}
\usepackage{graphicx}
\usepackage{fix-cm}
\def\ii{\'{\i}}

\begin{document}

\markboth{Manuel Asorey and Jos\'e M. Mu\~noz-Casta\~neda }
{Boundary Effects In Quantum Physics}

%
\catchline{}{}{}{}{}
%

\title{{\bf {BOUNDARY EFFECTS IN QUANTUM PHYSICS}}}

\author{\bf MANUEL ASOREY}

\address{ Departamento de F\ii sica Te\'orica, Facultad de Ciencias, Universidad de Zaragoza\\
E-50009 Zaragoza, Spain\\
asorey@unizar.es}

\author{\bf JOS\'E M.  MU\~NOZ-CASTA\~NEDA}

\address{ Institut f\"ur Theoretische Physik, Universit\"at Leipzig, \\
Leipzig, G-04103, Germany\\
jose.munoz-castaneda@uni-leipzig.de}

\maketitle

\begin{history}
\end{history}

\newcommand{\combi}[2]{\left(\begin{array}{c} #1\\ #2\end{array}\right)}
\newcommand{\ham}{\mathbb{H}}
\newcommand{\comp}{\mathbb{C}}
\newcommand{\kop}{\widehat{K}}
\newcommand{\npro}{\mathcal{N}}
\newcommand{\CE}{\mathcal{E}}
\newcommand{\mpro}{\mathcal{M}}
\newcommand{\cpro}{\mathcal{C}}
\newcommand{\upro}{\mathcal{U}}
\newcommand{\opro}{\mathcal{O}}
\newcommand{\dpro}{\mathcal{D}}
\newcommand{\hpro}{\mathcal{H}}
\newcommand{\vpro}{\mathcal{V}}
\newcommand{\I}{\mathbb{I}}
\newcommand{\Z}{\mathbb{Z}}
\newcommand{\R}{\mathbb{R}}
\newcommand{\difn}[1]{d#1^{^{n+1}}}
\newcommand{\difnes}[1]{d^{^{n}}\!\!#1}
\newcommand{\dift}[1]{\frac{d^{^{3}}\!#1}{(2\pi)^3}}
\newcommand{\escalar}{\partial^{2}+\frac{m^{2}}{\hbar^{2}}}
\newcommand{\escalarep}{\partial^{2}+\frac{m^{2}}{\hbar^{2}}-i\epsilon}
\newcommand{\factorp}{(2\pi\hbar)^{n+1}\delta^{^{(n+1)}}}
\newcommand{\xes}{\textrm{\textbf{x}}}
\newcommand{\yes}{\textrm{\textbf{y}}}
\newcommand{\difdes}[1]{d^{^{2}}\!\!#1}
\newcommand{\difet}[1]{d^{^{3}}\!\!#1}
\newcommand{\partialb}{\overline{\partial}}
\newcommand{\az}{A_z}
\newcommand{\azb}{A_{\overline{z}}}
\newcommand{\zb}{\overline{z}}
\newcommand{\Db}{D_{\overline{z}}}
\newcommand{\kes}{\textbf{k}}
\newcommand{\qes}{\textbf{q}}
\newcommand{\segvar}{\widehat{K}}
\newcommand{\proyec}{\widehat{\mathcal{P}}}
\newcommand{\ldos}[1]{{L}^2(#1)}
\newcommand{\lldos}[1]{{H}^2(#1)}
\newcommand{\rldos}[1]{{H}^{\frac{1}{2}}(#1)}
\newcommand{\rdos}[1]{{H}^{-\frac{1}{2}}(#1)}
\newcommand{\rrdos}[1]{{H}^{-\frac{3}{2}}(#1)}
\newcommand{\rrldos}[1]{{H}^{\frac{3}{2}}(#1)}
\newcommand{\wpsi}{\widetilde{\psi}}
\newcommand{\wphi}{\widetilde{\varphi}}
\newcommand{\rphi}{\underline{\varphi}}
\newcommand{\half}{\frac{1}{2}}
\newcommand{\dsl}{/\!\!\!\partial}
\newcommand{\Dsl}{/\!\!\!\!D}
\newcommand{\tr}{{\rm tr}}
\newcommand{\e}{{\rm e}}
\newcommand{\spec}{{\rm spec}}
\newcommand{\li}{{\rm li}}
\newcommand{\re}{{\rm Re}}
\newcommand{\im}{{\rm Im}}
\newcommand{\ce}{\rm Casimir energy}
\newcommand{\mproT}{\upro(\ldos{M}{\comp})}

\begin{abstract}

We analyze the role  of boundaries in the infrared behavior of quantum field theories.
By means of a novel method  we calculate  the vacuum energy for a massless scalar field
  confined between two homogeneous parallel plates  with the most general type of boundary properties. This allows the discrimination between boundary conditions which generate attractive or repulsive Casimir forces between the plates. In the interface   between  both regimes we find a   very interesting family of boundary conditions  which do not induce any type of Casimir force. We analyze the effect of the renormalization group flow on these boundary conditions. 
 Even if the  Casimirless conformal invariant conditions  are  physically unstable under renormalization group flow they emerge as a new set of conformally invariant boundary conditions which are anomaly free.

\end{abstract}

\section{Introduction}
\label{intro}

The  role of boundaries of quantum systems in new physical phenomena has been a focus of increasing activity   in different areas of physics. In general,  the presence of boundaries enhances quantum aspects of the system.  Boundary properties
play a relevant role in the double slit Young experiment, Aharonov-Bohm effect, Casimir effect \cite{casimir48}, appearance of edge states and quantization of conductivity 
in the quantum Hall effect, and 
graphene physics.
On the other hand, the physics of boundary conditions also reachs the very foundations 
of fundamental physics: topological fluctuations in quantum gravity, 
black hole quantum physics, 
quantum holography, 
string theories and D-branes,
AdS/CFT correspondence 
and M-theory. 
There are also implications in  cosmology, 
where the recently observed suppression and alignment of quadrupolar and octupolar terms, might be related to  boundary conditions or non-trivial topology of the universe.

Boundary phenomena determine the structure of the vacuum and the low energy behavior of the quantum field theories. In massless theories   these effects are amplified because  the existence of long distance correlations allows  boundary effects to percolate the whole interior region. In that case the vacuum energy is highly dependent on the geometry of the bodies and the physical properties of the  boundaries \cite{miloni94,grib,mostep,bordag01,milton,kmm,book2}.  

In this paper we shall  focus on  the dependence of vacuum energy on boundary conditions in a massless field theory on a
domain bounded by two homogeneous parallel plates. The behavior of this energy with the distance between the plates is 
the basis for the Casimir effect. The energy due to vacuum fluctuations of the quantum fields induces a force between the plates and this force is always attractive for  two identical plates   because of the Kenneth-Klich theorem \cite{KK}. This theorem  shows from very general principles  that the force induced by the quantum vacuum fluctuations  between  two identical bodies is always attractive. 
However, there is an enormous interest in getting physical configurations where the Casimir force is repulsive instead of attractive.
One reason for this interest are the  technical applications to 
micro-mechanical devices (MEMS).   Another reason is that the existence of repulsive or null Casimir forces allows a more accurate   analysis of micro-gravity effects, because of the tight competition    between gravity and Casimir force at short distances. In particular, they are useful to verify the recently formulated conjectures about the violation of Newton gravitational law at sub-millimeter scales \cite{hoyle}. 

All  methods used to achieve a repulsive Casimir effect are based on  plates with different
properties. In fact,  new repulsive regimes of the Casimir effect have been found between plates with  different dielectric properties \cite{capasso},   non-local boundary conditions \cite{se}
 and between a metallic plate with a hole and a needle pointing to  the hole center.
 In this paper we consider the most general boundary 
conditions for two plates to analyze in great detail the transition from the attractive Casimir regime to the repulsive Casimir regime \cite{brevik, tesis, gadella,orsay}.
We also analyze the boundary renormalization group flow of boundary conditions and the stability properties of the Casimirless
boundary conditions \cite{orsay}.
Although in practice, only some of these boundary conditions can be physically implemented, the advances on nano-science allows to construct new materials (metamaterials) with very special characteristics, which can allow in near future the implementation of new types of boundary conditions.

\section{Quantum Fields in Bounded Domains}
\label{sec:1}

Let us consider  a free complex scalar field $\phi$ confined in a domain between two parallel plates
$\Omega=\{{\bf x}\in \R^3; 0<x_3<L\}$. 
The quantum Hamiltonian  given by
\begin{equation}
  {H}={1\over 2} \int_\Omega d^{3}{\bf x} \left( |\pi({\bf x})|^2 + {\phi}^\ast ({\bf x}) ({ -\mathbf{\Delta}+m^2)}\,\phi({\bf x})  \right)
\end{equation}
corresponds to  an infinite number of decoupled harmonic oscillators associated to the Fourier modes of 
the operator $ -\mathbf{\Delta}+m^2$ acting on static fields confined in $\Omega$. The stability conditions 
require that the corresponding
oscillating frequencies have to be real and positive. This condition can  be fulfilled for any value of the mass
only if all eigenvalues of the Laplacian operator $-\Delta$  are real and nonnegative, i.e. $-\Delta$ is a
self-adjoint non-negative operator.
The positivity condition can be relaxed for a fixed geometry and given mass, but the independence on the
size of $\Omega$, e.g. in the large volume limit, becomes equivalent to mass independence. In all cases 
the consistency is guaranteed by the positivity of the self-adjoint extension of $-\Delta$.

Because of the homogeneity  of the plates the boundary conditions are translation invariant along the plates.
Thus, the most general homogeneous boundary condition which satisfies the selfadjointness  condition of  $-\Delta$
is given by \cite{aim}
\begin{equation}
 \varphi - i L\dot\varphi = U (\phi +iL \dot\varphi),
 \label{bc}
\end{equation}
where $U$ is any $2\times 2$ unitary matrix,  $\varphi$ is the boundary value of $\phi$ and $\dot\varphi$ the normal derivative of $\phi$
at the boundary $\partial\Omega$. The positivity condition of $-\Delta$ imposes further restrictions in $U$. In consequence, the
set of all  boundary conditions which preserve unitarity are given by (\ref{bc}) with unitary matrices $U$ whose eigenvalues, $\lambda=\e^{i\alpha}$ satisfy that $0\leq\alpha\leq\pi$ \cite{gadella,orsay}. In summary, $U$ has to be of the form
\begin{equation}
  U(\alpha,\beta,\theta,\varphi)=\e^{i\alpha}\left(\cos(\beta)\I +i\sin(\beta)\,{{\bf n}}\cdot\bm\sigma\right) \label{parametrization}\,
\end{equation}
with ${0\leq \alpha\leq 2\pi,\,-\pi/2\leq \beta\leq \pi/2\ {\rm and}\   0\leq\alpha\pm\beta\leq\pi,}$  where $\bm{\sigma}$ are the Pauli matrices and ${{\bf n}}$
an unitary vector
 of the $S^2$ sphere.

  Some  symmetries of the classical theory can be broken upon quantization by
quantum interactions. In the case of fields confined in bounded domains only the symmetries which
leave the boundary invariant can be preserved for some boundary conditions. 
However, in the case of scale invariance ($x\rightarrow x/\Lambda)$  the presence of the boundaries does not automatically imply the
breaking of the symmetry at the quantum level because the rescaling involved in the Wilson renormalization
group transformation restores the system back to the same boundary domain $\Omega$.
Now,  scale invariance in the massless quantum field theory 
can still be broken because not all boundary conditions preserve this symmetry.
 In fact, it has been shown in Refs. \cite{barna,irgac} that
the renormalization group acts on the space of boundary conditions according to
the flow 
\begin{equation}
 \Lambda\, U_\Lambda^\dagger \partial^{\phantom\dagger}_{\Lambda} U^{\phantom\dagger}_\Lambda=  U_t^\dagger \partial_{t} U_t=\frac12 \left(U^\dagger_\Lambda -U^{\phantom\dagger}_\Lambda\right),
 \label{RGf}
\end{equation}
where  $ \Lambda=\Lambda_0\, {\mathrm{e}}^t.$

The only boundary conditions which  preserve scale  invariance are the fixed points of the renormalization
group flow (\ref{RGf}), i.e. boundary conditions whose unitary operators $U$ are 
 Hermitian unitary  matrices $ U^\dagger=U=U^{-1}$ \cite{irgac}.

In the parametrization given by (\ref{parametrization}) 
the flow reads (see Fig. 1)
\begin{equation}
  \alpha'(\Lambda)+{1\over\Lambda}\sin(\alpha)\cos(\beta)=0;\ 
  \beta'(\Lambda)+{1\over\Lambda}\cos(\alpha)\sin(\beta)=0;\
 {{\bf n}}'=0,
  \end{equation}
  which defines a vector field that can be extended to the whole group $U(2)$.

\begin{figure}[htbp]
\centerline{\includegraphics[height=5.5cm]{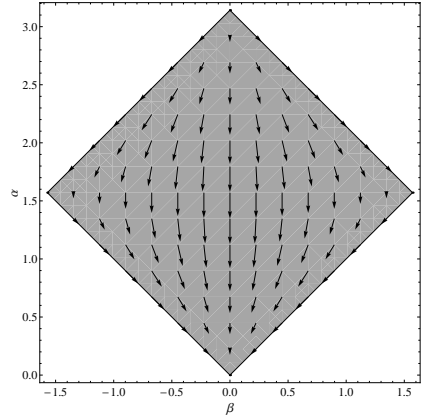}}
\caption{\footnotesize{Renormalization  group flow in the space of consistent boundary conditions. 
 The 
 box represents the projection on the plane $(\beta,\alpha)$ of the
 space of selfadjoint extensions of $-\Delta$, and the grey rhombus  the subset of non-negative selfadjoint 
 extensions.  Notice that fixed points  are located at the corners of the 
rhombus.  Neumann boundary conditions are at the lowest corner which is the  only stable fixed point.
This point is the final attractor of the whole renormalization group flow.}}\label{rgflux}
\end{figure}

All fixed points are located at the corners of the rhombus in figure \ref{rgflux}. The upper and
lower corners correspond to {Dirichlet  and Neumann}  ($ U=\mp \I $) boundary conditions.
The other fixed points are located at the other two corners and correspond to a $S^2$ manifold  
 given by
$
U={{\bf n}}\cdot\bm\sigma,
$
${\bf n}$ being an arbitrary unit vector of $\R^3$, which includes 
pseudo-periodic and  quasiperiodic  boundary conditions.

For mixed boundary conditions
$
 U=\mathrm{e^{2i \arctan e^{-t}}} {\mathrm {\I}}.
$
 the RG flows from   Dirichlet boundary conditions ({ultraviolet fixed point}) toward 
 Neumann boundary conditions ({infrared fixed point})  \cite{barna,irgac}.

\section{Vacuum energy.}
The Casimir effect in massless 	quantum theories is a consequence of  the scale symmetry anomaly 
which  arises in the form of finite size corrections to the vacuum energy.
Within the global framework of boundary conditions
formulated above it is possible to analyze with complete generality the characterization of attractive
and repulsive regimes generated by this anomaly.

The  vacuum energy 
 given by the sum  of the eigenvalues of $ \frac1{2} \sqrt{-\Delta_U}$ 
 is ultraviolet divergent, but the Casimir effect is associated to some finite volume corrections
 of the vacuum energy which are  UV finite and universal.
 In  a  heat kernel regularization  of UV divergences 
 \begin{equation}
{E^{(L,\epsilon)}_U}=\frac12\tr \sqrt{-\Delta_U} {\rm e}^{\epsilon {\Delta_U}},
\label{hk}
\end{equation}
 the Casimir energy  
 can be obtained from the asymptotic expansion in powers of $\frac{\sqrt{\epsilon}}{L}$ 
 of the vacuum energy per unit plate area $A$ \cite{valladolid},
\begin{equation}
{\frac{E^{(L,\epsilon)}_U}{A}=  \frac{c_0}{\epsilon^{{2}}} L+ \frac{c_1}{\epsilon^{3/2}} + \frac{c_{U}}{L^3} +
 \opro\left({\sqrt{\epsilon} \over L{}}\right)}.\label{asint}
\end{equation}

The
eigenvalues $\lambda_n=(k^1)^2+(k^2)^2+k_n^2$ of  the Laplacian operator $-\Delta_U$ are given in terms of  the zeros 
 $k_n$ of the spectral function \cite{tesis,gadella,orsay}
\begin{equation}
\begin{array}{lll}
\displaystyle  h_U(k)&=& 4 k e^{2i\alpha} \cos kL- 2 i  (1+k^2) e^{2i\alpha}\sin kL +  8 i  n_1 k  e^{i\alpha}\sin\beta 
  \phantom{\Bigr|}\cr
\displaystyle && -  2 i  (1+k^2) \sin kL -
 4 k \cos kL  +  4 i (1-k^2)e^{i\alpha} \cos\beta  \sin kL.\end{array}
\label{spf}
\end{equation}
and two arbitrary real parallel components $k^1,k^2$. The spectral function $h_U(k)$ is obtained  from
the determinant of the coefficients of the eigenvalue equation of $-\Delta_U$ 
for plane waves with momenta $(0,0,k)$.
The vacuum energy can be formally given in terms of the spectral function $h_U$ \cite{brevik,tesis,gadella}
(see \cite{mit, kirsten}) for an historical review) by
\begin{equation}
E_0= \frac{1}{24\pi^2 i}\oint {dz}{}\, z^3\,   \partial_z \log h_U(z)
= -\frac{1}{12 \pi^2}\int^\infty_0 {d}k \,  k^3\,   \frac{d}{dx}  \log h_U(i k).
\label{form}
\end{equation}
Using the heat kernel regularization (\ref{hk}) and the asymptotic expansion (\ref{asint})
 it is possible to compute  in a very efficient way from expression (\ref{form}) the Casimir energy for arbitrary  boundary conditions.

In some cases the Casimir energy can be computed analytically \cite{tesis,brevik,elizalde89,elizb,aa,elizalde03,gadella,orsay}.  
The results  summarized in table 1
\begin{table}[h]
\begin{center}
\begin{tabular}{|c|l|}
\hline
$\, U_n\,\raise2.5pt\hbox{{$_=$}}\, \raise2.5pt\hbox{{$_-$}}U_d\,\raise2.5pt\hbox{{$_=$}}\,\I \,  $&$  {c}^{}_{\mathrm{n}}\,\raise2.5pt\hbox{{$_=$}}\, {c}^{}_{\mathrm{d}}\,\raise2.5pt\hbox{{$_=$}}\,\raise2pt\hbox{{$_-$}}\frac{\pi^2}{1440}$ $\phantom{\bigg\} }$\\
\hline
$\,U_{{\mathrm{ap}}}\,\raise2.5pt\hbox{{$_=$}}\,\raise2.5pt\hbox{{$_-$}}\sigma_1\,  $&${c}^{}_{{\mathrm{ap}}}\,\raise2.5pt\hbox{{$_=$}}\,\frac{7 \pi^2}{720}$ $\phantom{\bigg\} }$\\
\hline
$\,U_{{\mathrm{p}}}\,\raise2.5pt\hbox{{$_=$}}\,\sigma_1\,  $&$  {c}^{}_{\mathrm{p}}=\raise2.5pt\hbox{{$_-$}}\frac{\pi^2}{90}  \phantom{\bigg\} }$\\
\hline
$\, U_{{\mathrm{z}}}\,\raise2.5pt\hbox{{$_=$}}\,\raise2.5pt\hbox{{$_\pm$}}\sigma_3\,  $&$  {c}^{}_{{\mathrm{z}}}=\frac{7 \pi^2}{11520 }$ $\phantom{\bigg\}} $\\ 
\hline
$\, U_{{\mathrm{qp}}}\,\raise2.5pt\hbox{{$_=$}}\,\cos\theta\,\sigma_3\raise2.5pt\hbox{{$_+$}}\sin\theta\, \sigma_1 \,  $&$ {c}^{}_{{\mathrm{qp}}}\,\raise2.5pt\hbox{{$_=$}}\,\!\frac{127\pi^2}{11520}\raise2.5pt\hbox{{$_-$}}\frac{3\pi\theta}{32}\raise2.5pt\hbox{{$_-$}}\frac{11\theta^2}{96}\raise2.5pt\hbox{{$_-$}} \frac{4\theta^3+ |{\pi}-2\theta|^3}{96\pi}\raise2.5pt\hbox{{$_+$}}\frac{\theta^4}{48\pi^2};\ {\textstyle\theta\in\left[0,\,{\pi}\right]}$  $\phantom{\bigg\}}$\!\!\!\!\!\!\\
\hline
$\, U_{{\mathrm{pp}}}\,\raise2.5pt\hbox{{$_=$}}\,\cos\xi\, \sigma_1\raise2.5pt\hbox{{$_-$}} \sin\xi\, \sigma_2\,  $&$  {c}^{}_{{\mathrm{pp}}}(\xi)\,\raise2.5pt\hbox{{$_=$}}\,\raise2.5pt\hbox{{$_-$}}\frac{\pi ^2}{90}+\frac{\xi^2}{12}\raise2.5pt\hbox{{$_-$}}\frac{\xi^3}{12 \pi }\raise2.5pt\hbox{{$_+$}}\frac{\xi^4}{48 \pi ^2};\quad {\textstyle \xi\in[0,\,2\pi]}$ 
$\phantom{\bigg\}}$\\
\hline
\end{tabular}
\end{center}
\caption{Casimir energies  for different boundary conditions obtained by different analytical methods: Dirichlet (d), Neumann (n),
antiperiodic (ap), periodic (p), Zaremba (z), quasiperiodic (qp) and pseudoperiodic (pp).}
\end{table}
show that many of the conditions (e.g. Dirichlet, Neumann, periodic) give rise to attractive forces between the 
plates, others (e.g. antiperiodic, Zaremba) induce  repulsive forces, and between these two types of boundary conditions there exist a family of boundary conditions with no Casimir force \cite{brevik}.
In the case of quasi-periodic boundary conditions for
  $\theta^{\pm}_{\mathrm{qp}}={\pi[{1\over2}\pm(1-\textstyle{({1-2\hbox{$\sqrt{{2/ 15}}$}})^\half)]}}$,
the Casimir energy vanishes which signals the transition from attractive to
repulsive regimes of the Casimir effect. Indeed,
something similar occurs in the case of  pseudo-periodic boundary conditions where there are two values of $\xi$ with vanishing Casimir energy, and therefore there is no force between plates
$  \xi^{\pm}_{\mathrm{pp}}=\textstyle{\pi[1\pm(1-2\hbox{$\sqrt{{2/15}}$})^\half]}.$

Another particular case of interest is the case of fixed points of the renormalization group which are saddle points  and are located at
 left and right corners of the rhombus of Fig. 1, i.e.  boundary conditions corresponding to  points on the unit sphere $S^2$ for values $\alpha=\pm\beta=\frac{\pi}{2}$. These include periodic, anti-periodic, quasi-periodic, and pseudo-periodic boundary conditions. The Casimir energy  given by
 \begin{equation}\textstyle{E_{{\mathrm{}}}(n_1)=\frac1{L^3}\left(-\frac{\pi ^2}{90}+\frac{(\arccos n_1)^2}{12}-\frac{(\arccos n_1)^3}{12 \pi }+\frac{(\arccos n_1)^4}{48 \pi ^2}\right)}.
\end{equation}
with $ \arccos\, n_1\!\!\in\!\![0,\,2\pi]$ has 
two  attractive and repulsive regimes separated by a one dimensional circle of Casimirless boundary conditions given
by $\textstyle {\alpha= \beta=\textstyle\frac{\pi}{2};  n_1=\cos\pi[\textstyle {1\pm({\textstyle 1-2\sqrt{{2/15}}})^\half}]}$.

The subspace of Casimirless boundary conditions is unstable under the renormalization group flow,
because  it only  intersects the 
manifold of fixed points at the $S^2$ sphere of saddle fixed points. 
Obviously, Dirichlet and  Neumann boundary conditions have always a non-vanishing atractive Casimir energy.

For more general boundary conditions it is possible  to numerically evaluate  the Casimir energy. In this way one can find the complete set of boundary conditions which give rise to attractive Casimir forces and those which give rise to repulsive forces \cite{gadella,orsay}. 
On the other hand, boundary conditions for identical plates correspond to $\beta=0$ and 
from the numerical calculations it is shown that all these boundary conditions are always in the attractive regime, 
which is in agreement with the Kenneth-Klich theorem \cite{KK}.

In summary, the global analysis of the dependence of  infrared properties of field theories  on the nature of boundary conditions unveils many interesting physical effects. However, the characteristics of  boundary conditions which encode the attractive or repulsive nature of  the Casimir energy are still unknown, although the algorithm found in the previous section provides the simplest mechanism to determine such a character. 
The powerful method based on the use of the spectral function for the calculation of the Casimir effect 
permits to analyze from a   global perspective the properties of the Casimir energy as a function over the space of consistent boundary conditions. 
On the other hand it will be very interesting  to understand the special role of the  Casimirless boundary conditions which are also fixed points of the boundary renormalization group. Even if these  Casimirless conformal invariant conditions  are  physically unstable under renormalization group flow they provide a new set of conformally invariant boundary conditions which are anomaly free.
The existence of similar conditions in 1+1 dimensions opens a new approach for the study of string theory in non-critical dimensions. The  role of such conformally invariant boundary conditions in the corresponding  string theory deserves further study.

\section*{Acknowledgments}
For many years one of us (M. A.) has been closely working with 
Beppe Marmo on  physics and mathematics of boundary conditions and
would like to thank him for his generous friendship much beyond
physical boundaries. Both of us acknowledge his crucial collaboration in many of
the results reported in the paper. This work has been partially supported by the Spanish CICYT grant FPA2009-09638 and DGIID-DGA (grant 2010-E24/2).

\end{document}